\documentclass[Journal,onecolumn]{IEEEtran}

\usepackage{fancyhdr}
\usepackage{cite}
\usepackage{caption}
\usepackage{graphicx}
\usepackage{mathtools}
\usepackage{psfrag}
\usepackage{subfigure}
\usepackage{url}
\usepackage{stfloats}
\usepackage{amsmath}
\usepackage{array}
\usepackage{fancyhdr}
\usepackage{epsfig}
\usepackage{amssymb}
\usepackage{color}
\usepackage{multirow}
\usepackage{amsmath,amssymb,amsfonts}
\usepackage{graphicx}
\usepackage{textcomp}
\usepackage{xcolor}
\usepackage{amsmath}
\usepackage{graphicx}
\usepackage{amssymb}
\usepackage{setspace}

\usepackage{amssymb}
\usepackage{setspace}
\usepackage{steinmetz}
\newcommand{\floor}[1]{\left\lfloor #1 \right\rfloor}

\begin{document}
%
\title{Cooperative Learning via Federated Distillation over Fading Channels}
%
%
%

\author{Jin-Hyun~Ahn$^{*}$,
        Osvaldo~Simeone$^{\dagger}$,
        and~Joonhyuk~Kang$^{*}$\\
        $^{*}$ Korea Advanced Institute of Science and Technology, School of Electrical Engineering, South Korea\\
        $^{\dagger}$ King’s College London, Department of Engineering, London, United Kingdom\\
        $^{*}$ wlsgus3396@kaist.ac.kr, jhkang@ee.kaist.ac.kr, $^{\dagger}$  osvaldo.simeone@kcl.ac.uk
        }

\maketitle


\begin{abstract}
Cooperative training methods for distributed machine learning are typically based on the exchange of local gradients or local model parameters. The latter approach is known as Federated Learning (FL). An alternative solution with reduced communication overhead, referred to as Federated Distillation (FD), was recently proposed that exchanges only averaged model outputs. While prior work studied implementations of FL over wireless fading channels, here we propose wireless protocols for FD and for an enhanced version thereof that leverages an offline communication phase to communicate ``mixed-up'' covariate vectors. The proposed implementations consist of different combinations of digital schemes based on separate source-channel coding and of over-the-air computing strategies based on analog joint source-channel coding. It is shown that the enhanced version FD has the potential to significantly outperform FL in the presence of limited spectral resources.
\end{abstract}

\section{Introduction}
Federated Learning (FL) adopts periodic exchanges of model weights between devices and a Parameter Server (PS) in order to improve the performance of locally trained machine learning models \cite{FL}. The problem of reducing the communication overhead of FL, e.g., via quantization, is an active area of study (see, e.g., \cite{HighDim}). An alternative solution to FL with reduced communication overhead, referred to as Federated Distillation (FD), was recently proposed in \cite{FD}. FD is inspired by classical work on distillation of machine learning models \cite{D1,D2,D3}, and it requires devices to exchange only average output vectors, rather than model weights, to be used as a regularizer for local training.
\begin{figure}[t]
    \centering
    \includegraphics[width=100mm]{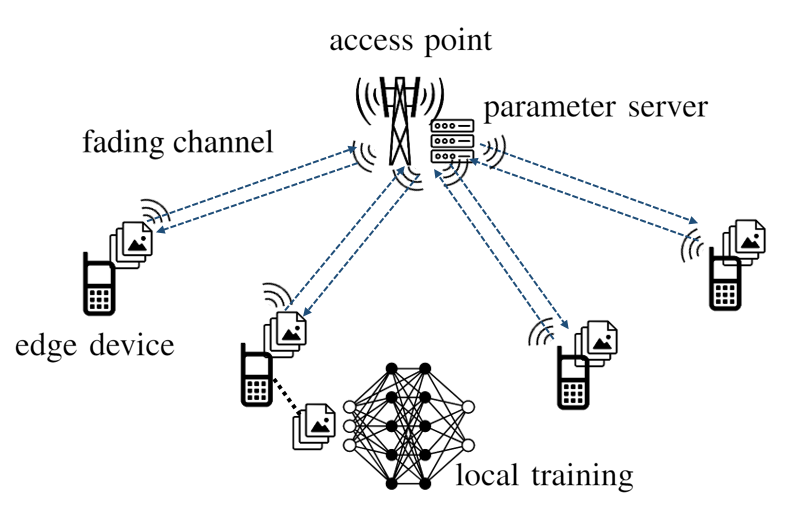}
        \caption{Edge training via wireless communications over fading channels through an access point.}
    \label{fig1}
\end{figure}

Implementing cooperative training schemes such as FL and FD over wireless channels requires the PS to compute the average of suitable local parameters. While this can be done using standard digital multiple access transmission schemes, recent work has leveraged the idea of over-the-air computing \cite{Nazer} in order to improve the efficiency in the use of spectral resources through analog transmission \cite{Deniz1,Deniz2,Low,PC2, PC1}. In particular, our previous paper \cite{HFD} proposed and analyzed implementations of FD, and of an enhanced version thereof termed Hybrid FD (HFD), over a Gaussian multiple access channel for the uplink and an ideal downlink channel. It is noted that HFD is closely related to the approach proposed more recently in \cite{FLD}, which is based on a combination of the mixup algorithm \cite{mixup} and FD.

In this work, we study the more challenging scenario in which the uplink is modelled as a multiple access fading channel and the downlink as a fading broadcast channel, as illustrated in Fig. 1. We develop implementations of FL, FD, and HFD that consist of different combinations of analog and digital strategies, and provide numerical comparisons.

\section{Problem Definition}\label{PD}
\subsection{System Set-Up}
As illustrated in Fig. \ref{fig1}, we consider a wireless edge learning system in which $K$ devices communicate via an Access Point (AP) over fading channels. Each device holds a local set $\mathbb{D}_{k}$ of data points. To enable cooperative training, the devices communicate over a shared fading channel with the AP, which is in turn connected to a Parameter Server (PS). The protocol prescribes a number of global iterations, with each iteration $i$ encompassing \textit{local training} at each device and \textit{information exchange} via the AP over the fading channels.

We focus on a classification problem with $L$ classes, with each dataset $\mathbb{D}_{k}$ consisting of pairs $\left(\mathbf{c}, \mathbf{t} \right) $, where $\mathbf{c}$ is the vector of covariates and $\mathbf{t}$ is the $L\times 1$ one-hot encoding vector of the corresponding label $t\in \left\{1, \dots, L \right\}$. Each device $k\in\left\{1,\dots,K\right\}$ runs a neural network model that produces the logit vector $\mathbf{s} \left(\mathbf{c}\vert \mathbf{w}^{k} \right)$ and the corresponding output probability vector $\hat{\mathbf{t}}\left(\mathbf{c}\vert \mathbf{w}^{k} \right)$ after the last, softmax, layer, for any input $\mathbf{c}$. The $W\times1$ weight vector $\mathbf{w}^{k}$ defines the network's operation at all layers. We recall that, for any given logit vector $\mathbf{s}=\left[s_1 , \dots, s_{L} \right]$, the output probability vector is given as 
\begin{equation}
    \hat{\mathbf{t}}\left(\mathbf{s}\right)= \left(\sum\limits_{i=1}^{L} e^{s_i}\right)^{-1}\begin{bmatrix} 
e^{s_1},\dots,e^{s_L}
\end{bmatrix}^{T},
\end{equation}
and we have $\hat{\mathbf{t}}\left(\mathbf{c}\vert \mathbf{w}^{k} \right)= \hat{\mathbf{t}}\left(\mathbf{s} \left(\mathbf{c}\vert \mathbf{w}^{k}\right) \right)$.

\subsection{Channel Model}
During each information exchange phase of the $i$-th global iteration, devices share a fading uplink multiple-access channel
\begin{equation}\label{receivedsignal}
\mathbf{y}_{i}=\sum_{k=1}^{K}  h^{k}_{i} \mathbf{x}^{k}_{i} + \mathbf{z}_{i},
\end{equation}
where $h^{k}_{i}$ is the quasi-static fading channel from the device $k$ to the AP; $\mathbf{x}^{k}_{i}$ is the $T_{U} \times 1 $ signal transmitted by the device $k$; and $\mathbf{z}_{i}$ is $T_{U}\times 1 $ noise vector with independent and identically distributed (i.i.d.) $\mathcal{CN} \left(0,1\right)$ entries. Each device $k$ has a power constraint $  \mathrm{E} \left[ \| \mathbf{x}^{k}_{i}\|^{2}_{2}\right] /T_{U} \leq P_{U}$. Furthermore, in each $i$-th global iteration, the AP can broadcast to all the devices in the downlink, so that the received signal from AP to device $k$ is 
\begin{equation}\label{receivedsignal2}
\mathbf{y}_{i}^{k}= g^{k}_{i} \mathbf{x}_{i} + \mathbf{z}_{i}^{k},
\end{equation}
where $\mathbf{x}_{i}$ is the $T_{D}\times 1 $ signal transmitted by the AP; $g^{k}_{i}$ is the quasi-static fading channel from the AP to the device $k$; and $\mathbf{z}_{i}^{k}$ is $T_{D}\times 1 $ noise vector with i.i.d. $\mathcal{CN} \left(0,1\right)$ entries. The AP has a power constraint $ \mathrm{E} \left[ \| \mathbf{x}_{i}\|^{2}_{2}\right] /T_{D} \leq P_{D}$.

\subsection{Training Protocols}\label{subsec:protocol}
In this section, we briefly review the training protocols that will be considered in this work (see \cite{HFD} for detailed algorithmic tables). Throughout, we define the cross entropy between probability vectors $\mathbf{a}$ and $\mathbf{b}$ as $\phi(\mathbf{a},\mathbf{b})=-\sum_{l=1}^{L} a_l \log b_l$. As a benchmark, with \textit{Independent Learning} (IL), each learning model at device $k$ is trained on the local training set $\mathbb{D}_{k}$ by using Stochastic Gradient Descent (SGD) with step size $\alpha >0$ on the cross-entropy loss (see, e.g., \cite{briefML}). With \textit{Federated Learning} (FL) \cite{FL}, at each global iteration $i$, each device $k$ follows IL within the local training phase, and then it transmits the update $\Delta \mathbf{w}_{i}^{k}$ of the local weight vector $\mathbf{w}^{k}_{i} $ to the PS during the information exchange phase. The PS computes the average update $\Delta \mathbf{w}_{i}= 1/K\sum_{k=1}^{K}\Delta\mathbf{w}^{k}_{i}$ with respect to the previous iteration. This is broadcast to all devices and used to update the initial weight vector for the local training phase in the next iteration. 

With \textit{Federated Distillation} (FD) \cite{FD}, each device $k$, during the information exchange phase of any iteration $i$, transmits the average logit vectors 
\begin{equation}\label{FDlocalavg}
    \mathbf{s}_{i,t}^{k}=\mathrm{E}_{\left(\mathbf{c},\mathbf{t}'\right)\in \mathbb{D}_{k}} \left[\mathbf{s}\left(\mathbf{c}\;\middle|\;\mathbf{w}^{k}_{i}\right)\;\middle|\; t'=t\right]
\end{equation}
for all labels $t=1,\dots,L$. In practice, the average in \eqref{FDlocalavg} is computed using a sample of data points from $\mathbb{D}_{k}$. The PS computes the average of the logit vectors, $ \mathbf{s}_{i,t}=1/K \sum_{k=1}^{K}\mathbf{s}_{i,t}^{k}$, which is transmitted to all devices in the downlink. During the local training phase of the next iteration $i+1$, given any selected data point $\left(\mathbf{c},\mathbf{t}\right)$, the training at each device $k$ is carried out via SGD with step size $\alpha > 0 $ on a regularized loss function. This is given by the weighted sum of the regular cross-entropy loss and of the cross-entropy $\phi(\hat{\mathbf{t}}(\mathbf{c}|\mathbf{w}^{k}_{i}),\hat{\mathbf{t}}(\mathbf{s}_{i,t}^{\backslash k }))$ between the local probability vector $\hat{\mathbf{t}}(\mathbf{c}|\mathbf{w}^{k}_{i})$ and the probability vector corresponding to the average logit vector for label $t$ (see [13, Eq. (7)]), i.e.,
\begin{equation}\label{FDbackslash}
    \mathbf{s}_{i,t}^{\backslash k }=\frac{K\mathbf{s}_{i,t}-\mathbf{s}_{i,t}^{k}}{K-1}.
\end{equation}

In HFD, which can be interpreted as a form of mixup \cite{mixup} (see also \cite{FLD}), during an additional offline phase, each device $k=1,\dots,K$ calculates the average covariate vectors $\tilde{\mathbf{c}}_{t}^{k}=1/\left|\mathbb{D}_{k}\right| \sum_{(\mathbf{c},\mathbf{t}) \in \mathbb{D}_{k}} \mathbf{c}$ for every label $t=1,\dots,L$ in the local dataset $\mathbb{D}_{k}$, which are uploaded to the PS. Then, the PS calculates the global average covariate vectors $\tilde{\mathbf{c}}_{t}=1/K\sum_{k'=1}^{K}\tilde{\mathbf{c}}_{t}^{k'}$ for all labels $t=1,\dots,L$. Finally, each device $k$ downloads $\tilde{\mathbf{c}}_{t}$ and calculates the vectors
\begin{equation}\label{HFDImagebackslash}
\tilde{\mathbf{c}}_{t}^{\backslash k}= \frac{K \tilde{\mathbf{c}}_{t}-\tilde{\mathbf{c}}_{t}^{k} }{K-1}
\end{equation}
for all labels $t=1,\dots,L$ in a manner similar to the logit vector \eqref{FDbackslash}.
At run time, during each local training phase, each device $k$ first carries out a number of SGD steps on the weighted sum of the regular cross-entropy loss and of the cross-entropy $\phi(\hat{\mathbf{t}}(\tilde{\mathbf{c}}_{t}^{\backslash k}|\mathbf{w}^{k}_{i}),\hat{\mathbf{t}}(\mathbf{s}_{i,t}^{\backslash k }))$ between the local probability vector $\hat{\mathbf{t}}(\tilde{\mathbf{c}}_{t}^{\backslash k}|\mathbf{w}^{k}_{i})$ and the probability vector corresponding to the average logit vector $\mathbf{s}_{i,t}^{\backslash k}=\mathbf{s}(\tilde{\mathbf{c}}_{t}^{\backslash k}|\mathbf{w}^{k}_{i})$ (see [13, Eq. (7)]). Then, each device performs a number of SGD updates following IL on the local dataset.

\section{Wireless Cooperative Training Over Fading Channels}\label{sec:wireless}
In this section, we propose wireless implementations for the cooperative training schemes summarized in Sec. \ref{subsec:protocol}.
Four implementations of the training protocols are proposed, which use either digital (D) or analog (A) communication in uplink and downlink. Accordingly, we distinguish among digital-digital (D-D), digital-analog (D-A), analog-digital (A-D), and analog-analog (A-A) protocols, with the two qualifiers referring to the uplink and downlink communications, respectively. Digital transmission for both uplink and downlink is based on separate source-channel coding \cite{Deniz1,Deniz2}, while analog transmission implements joint source-channel coding through over-the-air computing.

For future reference in this section, it is useful to define the following functions. The function $\mathrm{sparse}_{q} \left(\mathbf{u}\right)$ sets all elements of $\mathbf{u}$ to zero except for the largest $q$ elements and the smallest $q$ elements, which are dealt with as follows. Denoting the mean values of the remaining positive elements and negative elements respectively by $\mu^{+}$ and $\mu^{-}$, if $\mu^{+}>\left|\mu^{-} \right| $, the negative elements are set to zero and all the elements with positive values are set to $\mu^{+}$ and vice versa if $\left|\mu^{-} \right| > \mu^{+} $. The function $\mathrm{thresh}_{q} \left(\mathbf{u}\right)$ sets all elements of $\mathbf{u}$ to zero except for the $q$ elements with the largest absolute values. Finally, function $Q_{b} \left( \mathbf{u} \right)$ quantizes each non-zero element of input vector $\mathbf{u}$ using a uniform quantizer with $b$ bits per each non-zero element.

\subsection{Uplink Digital Transmission}
First, we introduce digital transmission for the uplink. While optimization of resource under digital communications was studied in \cite{Opdigital}, in this work, we consider for simplicity an equal resource allocation to devices as in \cite{Deniz1}. Accordingly, all $K$ devices share equally the number $T_{U}$ of channel uses \eqref{receivedsignal}, so that the number of bits that can be transmitted from each device $k$ per $i$-th global iteration is given as \cite{Thomas} 
\begin{equation}\label{Thomas}
    B_{U,k,i} = \frac{T_{U}}{K} \log_2 \left(1+ \left| h^{k}_{i} \right|^2 K P_{U} \right). 
\end{equation}
In order to enable transmission of the analog vectors required by FL, FD, and HFD, each device $k$ compresses the information to be sent to the AP to no more than $B_{U,k,i}$ bits at the $i$-th global iteration. Details for each learning protocol are provided next. Digital uplink schemes require each device $k$ to be aware of rate \eqref{Thomas}, and hence of the channel power $| h^{k}_{i} |^2$, and the AP to have full channel state information (CSI). 

\noindent \textbf{FL.} Under FL, each device $k$ at the $i$-th global iteration sends the $W \times 1$ update vector $\Delta \mathbf{w}_{i}^{k}$ to the AP. To this end, we adopt sparse binary compression with error accumulation  
\cite{Deniz1,DSGD}. Accordingly, each device $k$ at the $i$-th global iteration computes the vector $\mathbf{v}_{i}^{k}=\mathrm{sparse}_{q_{i}^{k}} \left( \Delta \mathbf{w}_{i}^{k} +\Delta_{i}^{k}\right)$, where the accumulated quantization error is updated as
\begin{equation}\label{errorupdate}
    \Delta_{i+1}^{k}=\Delta_{i}^{k} + \Delta \mathbf{w}_{i}^{k}- Q_{b} \left(\mathbf{v}_{i}^{k}\right).
\end{equation}
Then, it sends the $b$ bits obtained through the operation $Q_{b} \left(\mu \right)$, where $\mu$ is the non-zero element of $\mathbf{v}_{i}^{k}$, along with $\log_2 \binom{W}{q}$ bits specifying the indices of the $q$ non-zero elements in $\mathbf{v}_{i}^{k}$. The total number of bit to be sent by each device is hence given as $
       B^{FL}_{U,k,i} = b+ \log_2 \binom{W}{q_{i}^{k}},
$
where $q_{i}^{k}$ is chosen as the largest integer satisfying $B^{FL}_{U,k,i} \leq B_{U,k,i}$ for a given bit resolution $b$.


\noindent \textbf{FD and HFD.} Under FD and HFD, each device $k$ at the $i$-th global iteration should send the $L \times 1$ logit vector $\mathbf{s}_{i,t}^{k}$ in \eqref{FDlocalavg} for all labels $t= 1,\dots, L $. To this end, as in \cite{HFD}, each device $k$ computes the vector
$\mathbf{q}_{i,t}^{k}=Q_{b} (\mathrm{thresh}_{q_{i}^{k}} \left(\mathbf{s}_{i,t}^{k}\right))$, and the resulting bits are sent to the PS, along with the positions of the non-zero entries in vector $\mathbf{q}_{i,t}^{k}$ for all labels $t= 1,\dots, L $. The number of bits to be sent is hence given as $
    B^{FD}_{U,k,i} =L( b q_{i}^{k}+ \log_2 \binom{L}{q_{i}^{k}}),
$
where $q_{i}^{k}$ is chosen the largest integer satisfying $B^{FD}_{U,k,i} \leq B_{U,k,i}$. 


\subsection{Downlink Digital Transmission}
Under digital transmission in the downlink, the number of bits broadcast by the AP to all devices at the $i$-th global iteration is given as \cite{Thomas}
\begin{equation}
    B_{D,i} = \min_{k} \left(T_{D} \log_2 \left(1+ \left| g^{k}_{i} \right|^2 P_{D} \right)\right). 
\end{equation}
The PS compresses the information to be sent to the devices to no more than $B_{D,i}$ bits at the $i$-th global iteration. Downlink digital transmission requires the AP to have knowledge of the channel gain $\left| g^{k}_{i} \right|^2$ and each device $k$ to know the channel $g_{i}^{k}$.   

\noindent\textbf{FL.} The AP at the $i$-th global iteration sends the $W \times 1$ vector $\Delta \mathbf{w}_{i}$ obtained by averaging the decoded weight updates from the devices. As for the case of uplink, we adopt sparse binary compression with error accumulation. Therefore, the PS computes the vector $\mathbf{v}_{i}=\mathrm{sparse}_{q_{i}} \left( \Delta \mathbf{w}_{i} +\Delta_{i}\right)$, where the accumulated quantization error is updated as \eqref{errorupdate}. The total number of bit to send $Q_{b} \left(\mathbf{v}_{i}\right)$ is given as $B^{FL}_{D,i} = b+ \log_2 \binom{W}{q_{i}}$,
where $q_{i}$ is chosen as the largest integer satisfying $B^{FL}_{D,i} \leq B_{D,i}$. 

\noindent\textbf{FD and HFD.} Under FD and HFD, the AP at the $i$-th global iteration broadcasts the $L \times 1$ logit vector $\mathbf{s}_{i,t}$ obtained by averaging the decoded logit vectors from the devices for all labels $t= 1,\dots, L $. To send the quantized vector $\mathbf{q}_{i,t}=Q_{b} \left(\mathrm{thresh}_{q_{i}}\left(\mathbf{s}_{i,t}\right)\right)$, the number of bits is hence given as $
    B^{FD}_{D,i} =L ( b q_{i}+ \log_2 \binom{L}{q_{i}}),
$
where $q_{i}$ is chosen the largest integer satisfying $B^{FD}_{D,i} \leq B_{D,i}$. 



\subsection{Uplink Analog Transmission}\label{subsec:UA}
Under over-the-air computing, all the devices transmit their information simultaneously in an uncoded manner to the AP. The PS decodes the desired sum directly from the received signal \eqref{receivedsignal}. Different types of power allocation at the devices have been studied in the literature, namely full-power transmission, channel inversion \cite{Deniz2}, and optimized power control \cite{PC2, PC1}.
In this paper, full-power transmission is considered for simplicity, but extensions are conceptually straightforward. Since the vectors to be communicated in the uplink and downlink contain more samples than the number of available channel uses, these schemes generally rely on dimensionality reduction techniques, as detailed below for each protocol. Analog communication requires each device $k$ to have knowledge of the phase $\angle h_{i}^{k}$ of the channel $h_{i}^{k}$ to the AP, and the AP to know all channels.    

\noindent\textbf{FL.} In order to enable dimensionality reduction, assuming the inequality $T_{U}<W/2$, a pseudo-random matrix $\mathbf{A}_{U} \in \mathbb{R}^{ 2T_{U} \times W}$ with i.i.d. entries $\mathcal{N}(0,1/2T_{U})$ is generated and shared between the PS and the devices before the start of the protocol. In a manner similar to \cite{Deniz1,Deniz2}, each device $k$ at the $i$-th global iteration computes the sparsified vector $\mathbf{v}_{i}^{k}=\mathrm{thresh}_{q} \left( \Delta \mathbf{w}_{i}^{k} +\Delta_{i}^{k}\right)
$, for some $q$, where $\Delta_{i}^{k}$ denotes the accumulated error defined as \eqref{errorupdate}. To transmit the dimensionality-reduced vector $\hat{\mathbf{v}}_{i}^{k}=\mathbf{A}_{U} \mathbf{v}_{i}^{k}$, each device $k$ transmits vector $\mathbf{x}_{i}^{k} \in \mathbb{C}^{T_{U}\times 1}$, where
\begin{equation}\label{realtoimaginary}
    \mathbf{x}_{i}^{k} \left(m \right)=
    \hat{\mathbf{v}}_{i}^{k} \left(2m-1\right)+j\hat{\mathbf{v}}_{i}^{k} \left(2m\right),
\end{equation}
and $m=1,\dots,T_{U}$. By \eqref{realtoimaginary}, the transmitted signal encodes two different values of $\hat{\mathbf{v}}_{i}^{k}$ in the in-phase and quadrature components. Each device $k$ transmits the vector $\gamma_{i}^{k}  e^{-j \angle h_{i}^{k} }\mathbf{x}_{i}^{k} \in \mathbb{C}^{T_{U} \times 1}$, where the scaling factor $\gamma_{i}^{k}=  
    \sqrt{P_{U}T_{U}}/\|\mathbf{x}_{i}^{k} \|_{2} $ ensures full power transmission for the $k$-th device. The PS scales the received signal \eqref{receivedsignal} by the factor
\begin{equation}\label{rxscale}
    \nu_{i}= \frac{\sum\limits_{k'=1}^{K} \gamma_{i}^{k'} \left| h_{i}^{k'}\right| }{\frac{1}{2}+\sum\limits_{k'=1}^{K} \left(\gamma_{i}^{k'} \left| h_{i}^{k'}\right|\right)^{2} }
\end{equation}
in order to obtain a minimum mean square error estimate of the sum $ \mathbf{A}_{U}\sum_{k=1}^{K}\mathbf{v}_{i}^{k}$ \cite{PC2}. Finally, the PS applies a compressive sensing decoder such as Lasso or AMP \cite{Lasso,AMP} to this vector in order to estimate $\sum_{k=1}^{K}\mathbf{v}_{i}^{k}$.

\noindent\textbf{FD and HFD.} Under FD and HFD, each device $k$ at the $i$-th global iteration communicates the $L \times 1$ logit vector $\mathbf{s}_{i,t}^{k}$ for all labels $t= 1,\dots, L$. We assume here that the number $2T_{U}$ of real channel uses for communication slot is larger than $L^2$, since the number $L$ of classes is typically small. Otherwise, a dimension reduction scheme as described above could be readily used. Therefore, we can define the source integer bandwidth expansion 
factor $\rho=\floor{2T_{U}/L^{2}} \geq 1$. Under this condition, each device $k$ at the $i$-th global iteration implements $\rho$-fold repetition coding by transmitting $\mathbf{R}_{\rho}\mathbf{s}_{i}^{k}$, 
where matrix $\mathbf{R}_{\rho}=\mathbf{1}_{\rho} \otimes \mathbf{I}_{L^2}$, with $\mathbf{1}_{\rho}=(1,\dots,1)^{T}$, implements repetition coding with redundancy $\rho$; $\mathbf{I}_{L^2}$ is a $L^2 \times L^2$ identity matrix; and we have $\mathbf{s}_{i}^{k} = \left[(\mathbf{s}_{i,1}^{k})^{T} , \dots, (\mathbf{s}_{i,L}^{k})^{T} \right]^{T}$. 
To transmit the encoded vector $\mathbf{v}_{i}^{k}=\mathbf{R}_{\rho}\mathbf{s}_{i}^{k}\in \mathbb{R}^{\rho L^{2} \times 1}$, each device $k$ transmits $\gamma_{i}^{k}  e^{-j \angle h_{i}^{k} }\mathbf{x}_{i}^{k} \in \mathbb{C}^{\rho L^{2}/2 \times 1}$ where $ \mathbf{x}_{i}^{k} \in \mathbb{C}^{\rho L^{2}/2 \times 1}$ is defined as \eqref{realtoimaginary}. The PS scales the received signal \eqref{receivedsignal} by the factor \eqref{rxscale} and multiplies it by $\mathbf{R}_{\rho}^{T}/\rho$ to obtain an estimate of $\sum_{k=1}^{K}\mathbf{v}_{i}^{k}$.

\subsection{Downlink Analog Transmission}
 For the downlink broadcast communication from AP to devices, the AP transmits with full power and each device applies a scaling factor in order to estimate the vector transmitted by the AP, in a similar manner to analog transmission at the uplink. Details for each protocol are provided next.    

\noindent\textbf{FL.} In order to enable dimension reduction, a pseudo-random matrix $\mathbf{A}_{D} \in \mathbb{R}^{ 2T_{D} \times W}$ with i.i.d. entries $N(0,1/2T_{D})$ is generated and shared between the PS and the devices before the start of the protocol. At the $i$-th global iteration, the PS computes the sparsified vector $\mathbf{v}_{i}=\mathrm{thresh}_{q} \left( \Delta \mathbf{w}_{i} +\Delta_{i} \right)
$. To transmit the dimension-reduced vector $\hat{\mathbf{v}}_{i}=\mathbf{A}_{D} \mathbf{v}_{i}$, the AP transmits the vector $\gamma_{i}  \mathbf{x}_{i}$, where $\gamma_{i}=  
    \sqrt{P_{D}T_{D}}/\|\mathbf{x}_{i} \|_{2} $ ensures full power transmission and $\mathbf{x}_{i} \in  \mathbb{C}^{T_{D} \times 1 }$ is defined as \eqref{realtoimaginary}.
Each device $k$ scales the received signal \eqref{receivedsignal2} by scaling factor \cite{PC2}
\begin{equation}\label{rxscale2}
    \nu_{i}^{k}= \frac{\gamma_{i} \left| g_{i}^{k}\right| }{\frac{1}{2}+ \left(\gamma_{i} \left| g_{i}^{k}\right|\right)^{2} }.
\end{equation}
Finally, each device applies a compressive sensing decoder such as Lasso or AMP \cite{Lasso,AMP} to this vector in order to estimate $\mathbf{v}_{i}$. 

\noindent\textbf{FD and HFD.} 
Under FD and HFD, the PS at the $i$-th global iteration broadcasts the $L \times 1$ logit vector $\mathbf{s}_{i,t}$ for all labels $t= 1,\dots, L$. Similar to the case of uplink, we adopt the repetition coding with redundancy $\rho=\floor{2T_{D}/L^{2}} \geq 1$ and the AP transmits $\mathbf{v}_{i}=\mathbf{R}_{\rho}\mathbf{s}_{i}\in \mathbb{R}^{\rho L^{2} \times 1}$, where $\mathbf{s}_{i} = \left[(\mathbf{s}_{i,1})^{T} , \dots, (\mathbf{s}_{i,L})^{T} \right]^{T}$. The AP transmits $\gamma_{i} \mathbf{x}_{i} \in \mathbb{C}^{\rho L^{2}/2 \times 1}$ where $ \mathbf{x}_{i} \in \mathbb{C}^{\rho L^{2}/2 \times 1}$ is defined as \eqref{realtoimaginary}. Each device scales the received signal \eqref{receivedsignal2} by the factor \eqref{rxscale2} and multiply $\mathbf{R}_{\rho}^{T}/\rho$ to an estimated vector of $\mathbf{v}_{i}$.


\begin{figure}[b]
    \centering
      \includegraphics[width=100mm]{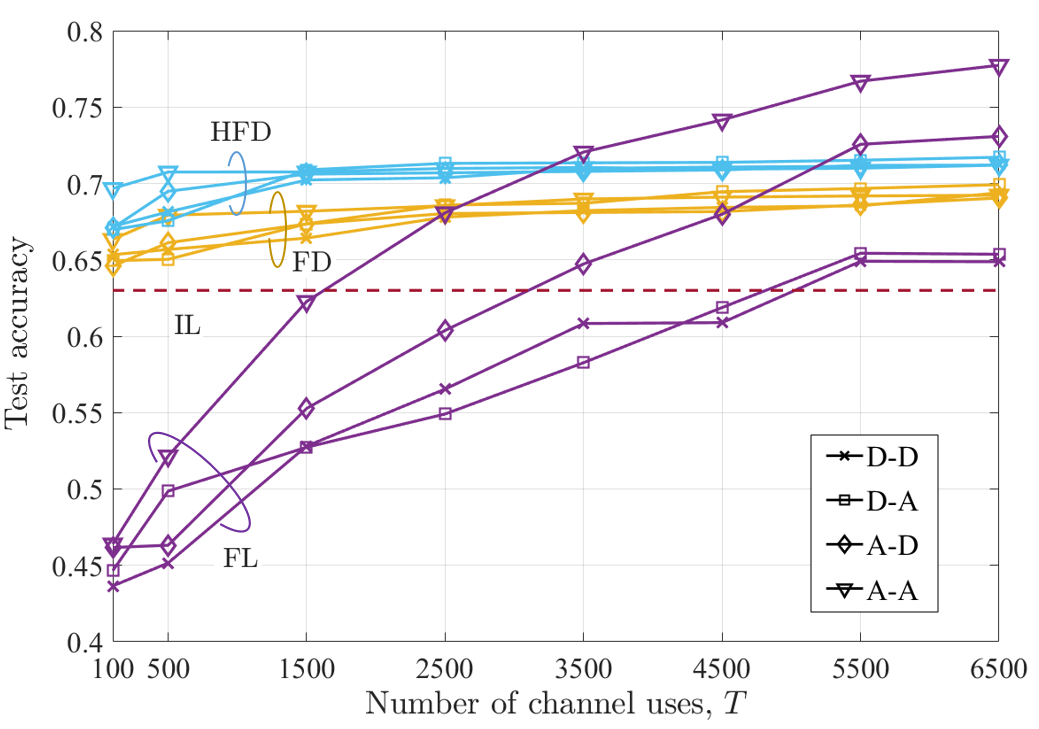}
     \caption{Classification test accuracy for IL, FL, FD, and HFD under implementations D-D, D-A, A-D, and A-A (A=analog, D=digital; first letter for uplink and second for downlink).}
    \label{fig2}
\end{figure}

\begin{figure}[t]
    \centering
      \includegraphics[width=100mm]{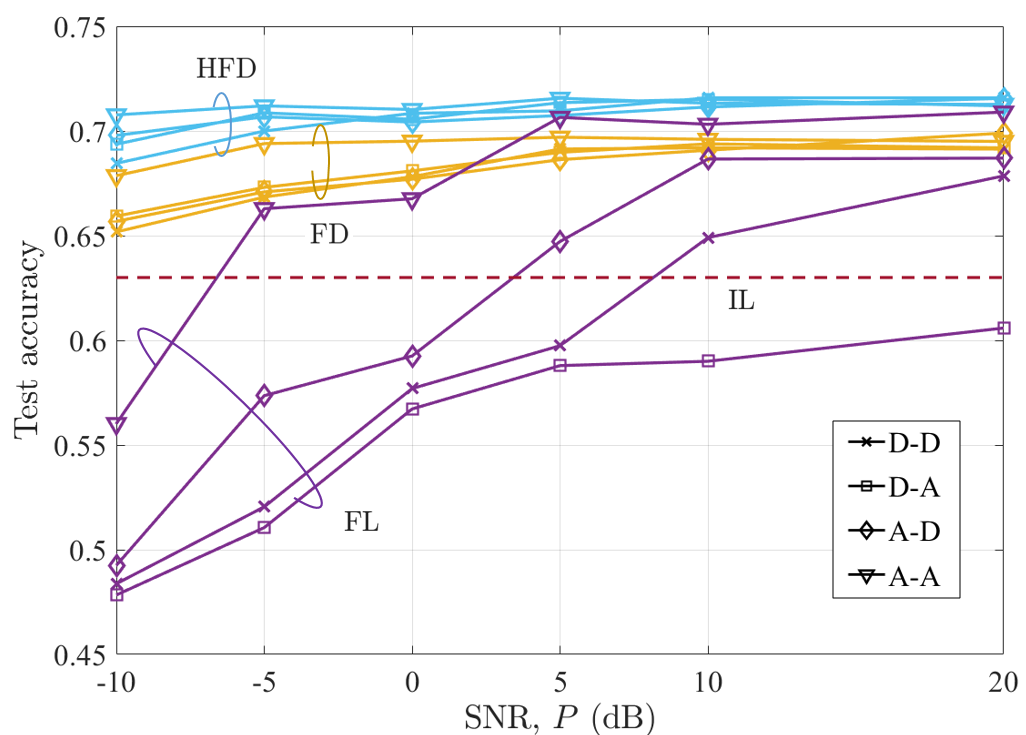}
     \caption{Classification test accuracy for IL, FL, FD, and HFD under implementations D-D, D-A, A-D, and A-A (A=analog, D=digital; first letter for uplink and second for downlink).}
    \label{fig3}
\end{figure}

\section{Numerical results and final remarks}\label{NR}
In this section, we consider an example with $K=10$ devices, each running a six-layer Convolutional Neural Network (CNN) that consists of two convolutional layers, two max-pooling layer, two fully-connected layers, and softmax layer to carry out image classification based on subsets of the MNIST dataset. Specifically, we randomly select disjoint sets of $64$ samples from the $60,000$ training MNIST examples, and allocate each set to a device. Note that, as a result, each device generally has unbalanced data sets with respect to the ten classes in the MNIST data set. We set to $10$ the number of global iteration; the SGD step size to $\alpha=0.001$; the number of quantization bits to $b=16$; the threshold level for analog implementation of FL to $q=4T/5$; and the number of uplink and downlink channel uses to $T_{U}=T_{D}=T$. 

The performance metric is the average test accuracy for all devices measured over $10,000$ randomly selected images from the MNIST dataset. In Fig. 2 and Fig. 3, the mentioned average test accuracy under IL, FL, FD, and HFD is plotted for the D-D, D-A, A-D, and A-A protocols introduced in Sec. \ref{sec:wireless}. In Fig. 2, the number $T$ of channel uses increases from $100$ to $6500$ while the signal-to-noise ratio (SNR) in the uplink is $P_{U}=0$ dB and the SNR in the downlink is $P_{D}=10$ dB. The key observation in Fig. 2 is that FD and HFD significantly outperform FL at low values of $T$, that is, with limited spectral resources. Furthermore, HFD is seen to uniformly improve over FD. For the implementations of FL, it is observed that the A-A scheme is clearly preferable over the alternatives. All implementations 
yield a similar test accuracy for FD and HFD due to their lower communication overhead, although the A-A scheme is still preferable at low values of $T$. 

In Fig. 3, the SNR in the uplink $P_{U}$ increases from $-10$ dB to $20$ dB while the SNR in the uplink is $P_{D}=P_{D}+10$ dB and the number $T$ of channel uses is $2500$. The figure confirms that FD and HFD significantly outperform FL at low values of $P$, and that HFD uniformly improves over FD. Furthermore, the A-A scheme shows the best performance, especially for lower values of $P$.

\section*{Acknowledgments}
The work of J. Ahn and J. Kang was supported by the National Research Foundation of Korea (NRF) grant funded by the Korea government (MSIT) (No. 2017R1A2B2012698). The work of O. Simeone was supported by the European Research Council (ERC) under the European Union's Horizon 2020 research and innovation programme (grant agreement No. 725731).

\end{document}